\documentclass[11pt,a4paper]{article}
\usepackage{color,helvet,times}
\usepackage{amsmath,lastpage,bm,textcomp}
\usepackage[numbers,sort&compress]{natbib}
\usepackage{csquotes}
\usepackage{graphicx}
\usepackage[T1]{fontenc}
\usepackage{helvet}
\usepackage[textwidth=160mm,textheight=250mm]{geometry}
\usepackage[width=0.8\textwidth]{caption}
\usepackage{fleqn}
\usepackage{color}
\usepackage{multirow}
\usepackage{multibbl}
\usepackage[open, openlevel=0]{bookmark}
\usepackage[onehalfspacing]{setspace}
\usepackage{bm}
\usepackage{float}
\usepackage{bibunits}

\usepackage{dcolumn}
\usepackage{bm}

\usepackage{amsmath}
\usepackage{newfloat}
\usepackage{amssymb}
\usepackage{amsfonts} 
\usepackage{wasysym}  
\usepackage{graphics}  
\usepackage{psfrag} 
\usepackage{graphicx} 
\usepackage{epsfig}    
\usepackage{rotating}  
 \usepackage[latin1]{inputenc}
 \usepackage{color}
 \usepackage{rotating}
\usepackage{csquotes}
 \usepackage{multirow}
\usepackage[nooneline, tight,raggedright,FIGTOPCAP]{subfigure}
\usepackage{hyperref}
\usepackage{epsfig}
\usepackage{placeins}

\DeclareFloatingEnvironment[name={FIGURE}]{suppfigure}
\newcounter{sequation}

\newcommand{\tn}[1]{\mathrm{#1}}
\newcommand{\avg}[1]{\langle #1\rangle} 

\newbibliography{main}
\newbibliography{supp}

\begin{document}
\begin{bibunit}
\begin{center}
\textbf{\huge{Quantifying protein diffusion and capture on filaments}}
\vspace{12pt}

\text{Emanuel Reithmann, Louis Reese, and Erwin Frey (\href{mailto:frey@lmu.de}{\nolinkurl{frey@lmu.de}})}
\vspace{12pt}

\textit{Arnold Sommerfeld Center for Theoretical Physics (ASC) and Center for NanoScience (CeNS), Department of Physics, Ludwig-Maximilians-Universit\"{a}t M\"{u}nchen, Theresienstra\ss e 37, D-80333 Munich, Germany, and Nanosystems Initiative Munich (NIM), Ludwig-Maximilians-Universit\"{a}t M\"{u}nchen, Schellingstra{\ss}e 4, D-80333 Munich, Germany}

\end{center}

\onehalfspacing
\begin{abstract}
{The functional relevance of regulating proteins is often limited to specific binding sites such as the ends of microtubules or actin-filaments. A localization of proteins on these functional sites is of great importance.
We present a quantitative theory for a diffusion and capture process, where proteins diffuse on a filament and stop diffusing when reaching the filament's end. 
It is found that end-association after one-dimensional diffusion is the main source for tip-localization of such proteins.
As a consequence, diffusion and capture is highly efficient in enhancing the reaction velocity of enzymatic reactions, where proteins and filament ends are to each other as enzyme and substrate. We show that the reaction velocity can effectively be described within a Michaelis-Menten framework.
Together one-dimensional diffusion and capture beats the (three-dimensional) Smoluchowski diffusion limit for the rate of protein association to filament ends. 
}
\end{abstract}
\vspace*{2.7pt}
\linespread{1.5}
The catalytic activity of enzymes is often restricted to specific binding sites.
The ends of microtubules (MTs) for example are binding sites for a plethora of MT associated proteins (MAPs)~\cite{Akhmanova2008}. At MT ends, MAPs can catalyze biochemical processes~\cite{Howard2007}, or serve as substrates for other enzymes. This makes an efficient association of MAPs to MT tips important.
Recent experiments suggest that one-dimensional diffusion of MAPs on MTs facilitates tip-targeting~\cite{Helenius2006,Cooper2009}. 
This idea goes back to the concept of \enquote{reduction in dimensionality} suggested by Adam and Delbr\"uck ~\cite{Adam1968} and has been largely applied~\cite{Hippel1989,Mirny2009}. 
However, a quantitative understanding of tip-binding mediated by diffusion on the filament and subsequent capture at the tip has remained elusive~\cite{Helenius2006,Cooper2010,Brouhard2008,Widlund2011,Powers2009,Dixit2009,Forth2014,Noujaim2014,Hansen2010,Klein2005}. 

Here we show that capturing at the tip is crucial for tip-localization of proteins. 
We present a theory where diffusion and capture is accurately quantified with an effective association rate constant and provide a result which depends only on experimentally accessible parameters. 
For proteins which are enzymatically active at filament ends, our theory predicts that diffusion and capture leads to an enhancement of the enzymatic reaction velocity due to stronger tip-localization. We observe that the reaction velocity in dependence of the enzyme concentration closely follows a Michaelis-Menten curve and quantify the contribution of one-dimensional diffusion to tip-localization and enzymatic processes downstream thereof.

\begin{figure}
\centering{\includegraphics[width=0.8\columnwidth]{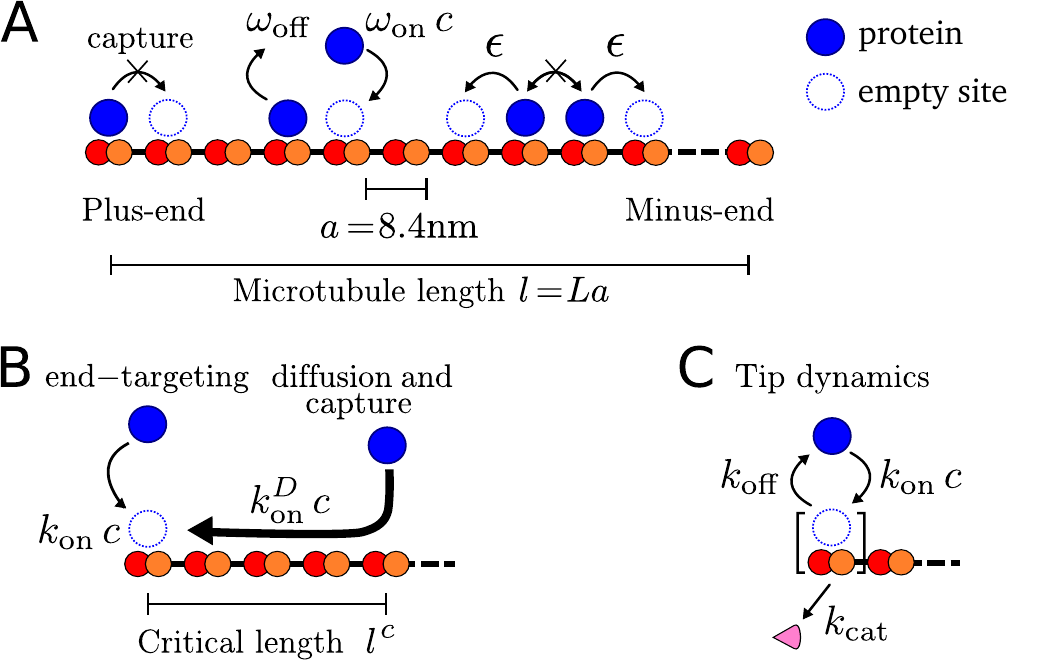}}
\caption{(A) Schematic of a MT with diffusive tip-binding proteins. In the bulk of the lattice, proteins attach to empty sites and detach. Proteins hop to neighboring sites but obey exclusion. At the plus-end, particles are captured. (B) Illustration of direct tip-attachment from solution and via diffusion and capture. (C) Proteins bind reversibly at the plus-end. While a protein is attached there, a reaction is catalyzed at rate $k_\tn{cat}$.  
\label{fig:model}}
\end{figure}

To model the diffusive motion of proteins on a filament we consider a one-dimensional lattice of length $l$ with lattice spacing $a = 8.4\,\tn{nm}$ [Fig.~\ref{fig:model}A].
The lattice corresponds to a single protofilament of a stabilized MT in the absence of dynamic instability. 
Proteins perform a random walk on the lattice with a hopping rate $\epsilon$; the diffusion constant is $D=\epsilon \, a^2$. Each site can be occupied by only one protein; the system is an exclusion process~\cite{Krapivsky2010}. 
Proteins attach to and detach from the lattice at rates $\omega_\tn{on} c$ and $\omega_\tn{off}$, respectively, where $c$ is the concentration of proteins in solution.
The tip of the MT is represented by the first lattice site in our model. To account for its particular structure, different on- and off-rates are assumed there, $k_\tn{on} c$ and $k_\tn{off}$. 
Proteins that bind to the tip are \emph{captured}, i.e. not allowed to hop on the lattice, but still may detach into solution. 
This important condition is a critical difference between our model and previous approaches~\cite{Helenius2006,Klein2005}, see also the Supporting Material.

\begin{figure}
\centering{\includegraphics[width=0.7\columnwidth]{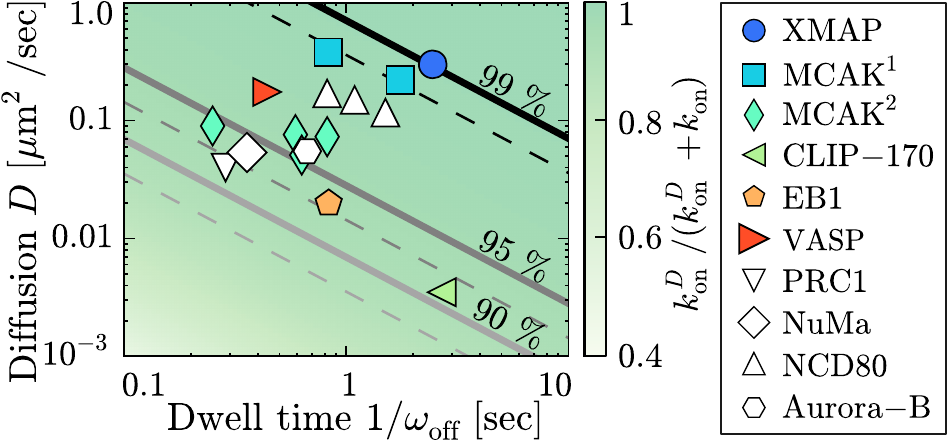}}
\caption{The model predicts the relative contribution to tip localization of proteins due to diffusion and diffusion \& capture (color code and solid lines), $k_\mathrm{on}^D/(k_\mathrm{on}^D+k_\mathrm{on})$ with $k_\mathrm{on}=\omega_\mathrm{on}$ (dashed for actin: $a=6\,\tn{nm}$). 
Proteins that are captured at the filament end (filled symbols) and proteins where evidence for capturing is lacking (open symbols) are shown. 
Proteins that in addition have a direct enzymatic activity at the filament end are $\mathrm{XMAP215}$~\cite{Brouhard2008,Widlund2011}, $\mathrm{MCAK}^1$~\cite{Helenius2006}, and $\mathrm{MCAK}^2$~\cite{Cooper2010} on MTs, and  $\mathrm{VASP}$ on actin filaments, see~Ref.~\cite{Hansen2010} and personal communicaton [S.D. Hansen and R.D. Mullins, 2014].
There are also proteins that diffuse on MTs without enzymatic activity at MT ends, but with roles downstream of tip-localization, e.g. in the protein network of MT tips~\cite{Akhmanova2008}: $\mathrm{NCD80}$~\cite{Powers2009}; $\mathrm{CLIP-170}$~\cite{Dixit2009}; $\mathrm{NuMa}$, $\mathrm{PRC1}$, $\mathrm{EB1}$~\cite{Forth2014}; $\mathrm{Aurora-B}$~\cite{Noujaim2014}. 
\label{fig:class}}
\end{figure}

A central goal of this letter is to quantify the relative contributions of \emph{diffusion and capture} (tip-attachment after diffusion on the lattice) and \emph{end-targeting} (attachment after diffusion in solution) [Fig.~\ref{fig:model}B] to tip-localization. To this end we calculated the probability to find a protein at the end of a protofilament (the tip density $\rho_+$). In the absence of diffusion and capture, the Langmuir isotherm is obtained, 
\begin{equation}
\rho_+ (c) = \frac{c}{K + c}  \, , \label{eq:rhoplus}
\end{equation}
where $K=k_\tn{off}/k_\tn{on}$ is the dissociation constant of the protein at the tip. However, as noted {pre\-viously}~\cite{Helenius2006,Cooper2009}, such a model is incomplete as it does not account for the additional protein flux along protofilaments mediated by diffusion and capture.
We have analyzed this flux by stochastic simulations of the model~[Fig.~\ref{fig:model}]. Surprisingly, we find that over a broad range of concentrations $c$, the additional protein current to an unoccupied reaction site $J^D$ effectively obeys first order kinetics, i.e. $J^D=k_\tn{on}^D c$ [Supporting Material Fig.~\ref{fig:linearity}]. 
This observation implies that despite the complexity of the diffusion-reaction process one approximately retains the functional form of the Langmuir isotherm. 
Accounting for the diffusion-capture contribution to the rate of protein attachment leads to an effective dissociation constant
\begin{equation}
K^\tn{eff}={k_\tn{off}}/{(k_\tn{on} + k_\tn{on}^D)}\, .\label{eq:Kmdiff}
\end{equation} 
We have calculated the diffusion-capture rate $k_\tn{on}^D$ analytically, by exploiting the observed approximate linear reaction kinetics. 
We find
\begin{equation}
k_\tn{on}^D 
= \frac{\omega_\tn{on} D / a^2}
       {\omega_\tn{off} 
+ \sqrt{\omega_\tn{off} D / a^2}}\,.\label{eq:kondiff}
\end{equation}
Refer to the Supporting Material for a detailed derivation of Eqs.~\ref{eq:rhoplus}-\ref{eq:kondiff}. Together Eqs.~\ref{eq:rhoplus}-\ref{eq:kondiff} comprise an effective theory for the association of proteins to the tip which accounts for direct end-targeting as well as the diffusion-capture process.
With Eq.~\ref{eq:kondiff} we are able to quantitatively predict the relative contribution of diffusion and capture to tip-binding for different proteins that diffuse on filaments. The results are shown in Fig.~\ref{fig:class}: $90-99\%$ of molecules bind to the tip through one-dimensional diffusion given they follow diffusion and capture. 

Tip-localization due to diffusion and capture as predicted by our theory has important implications for enzymatically active proteins. We extended the model to investigate enzymatic reactions at the MT tip, where
the protein-tip complex catalyzes a product at rate $k_\tn{cat}$ [Fig.~\ref{fig:model}C]. In detail, we assume that the protein does not leave the tip after catalyzing a reaction, but only through detachment into solution. 
These model assumptions are consistent with filament polymerizing enzymes that act processively, such as XMAP215 for MTs~\cite{Brouhard2008,Widlund2011}, and VASP~\cite{Hansen2010} and formins~\cite{Vavylonis2006} for actin filaments. The assumption of a constant length $l$ in our model is excellent if the rate of diffusion is fast compared to the polymerization rate.
With the above model assumptions the reaction velocity $v$ is determined by the tip density, $v = \rho_+ k_\mathrm{cat}$. We can apply our previous results, Eqs.~\ref{eq:rhoplus} -\ref{eq:kondiff}, to obtain
\begin{equation}
v(c)=k_\tn{cat}\, \rho_+(c) = \frac{k_\tn{cat} c}{K^\tn{eff} + c}\, .
\label{eq:speed}
\end{equation}
The above equation is reminiscent of a single-molecule Michaelis-Menten equation~\cite{Kou2005, Michaelis1913} when $K^\mathrm{eff}$ is reinterpreted as Michaelis constant and substrate and enzyme concentrations are interchanged. In this way our theory constitutes an effective Michaelis-Menten theory which accounts for end-targeting \emph{and} diffusion and capture; instead of solving a complex many-body problem it suffices to apply a mathematical framework which is analogous to (single-molecule) Michaelis-Menten kinetics; the details of diffusion and caputure are accurately included in the effective on-rate $k_\mathrm{on}^\mathrm{eff} = k_\mathrm{on} + k_\mathrm{on}^D$. This result is in accordance with experimental results for several enzymatically active proteins where Michaelis-Menten curves were observed for the reaction speed depending on the enzyme concentration~\cite{Brouhard2008,Cooper2010}.
Inspired by the processive (de)polymerase activity of (MCAK) XMAP215, we assume that enzyme and substrate are not decomposed in the reaction step. However, it is straightforward to include a decomposition in the theory: the effective dissociation constant would read $K^\mathrm{eff} ={(k_\tn{off}+k_\tn{cat})}/ k_\mathrm{on}^\mathrm{eff}$.  

\begin{figure}[t]
\centering{\includegraphics[width=0.6\columnwidth]{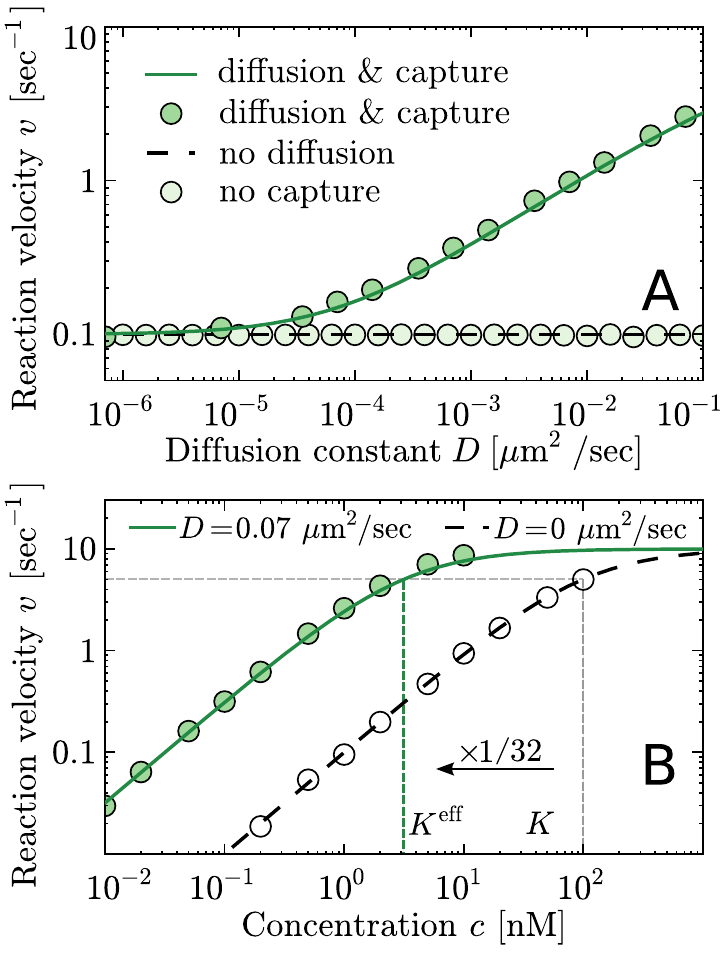}}
\caption{(A) Comparison of the reaction velocity with (solid) and without (dashed) lattice diffusion and with and without capturing at the tip (circles: simulation data; lines: analytic results) (B) shows the reaction velocity $v(c)$. Analytic results (lines) are confirmed by simulation data (circles). Parameters are $L=1000$, $\omega_\mathrm{off}=k_\mathrm{off}=1\, \mathrm{sec}^{-1}$, $k_\mathrm{cat}=10\, \mathrm{sec}^{-1}\, \omega_\mathrm{on}=k_\mathrm{on}= 0.01 \, \mathrm{sec}^{-1}\, \mathrm{nM}^{-1}$ and $c= 1\,\mathrm{nM}$. 
\label{fig:velocity}
}
\end{figure}

Our analytical results, Eqs.~\ref{eq:Kmdiff}-\ref{eq:speed}, agree well with simulation results of the stochastic model, as shown in Fig.~\ref{fig:velocity}(A) and (B). 
We find that the diffusion and capture mechanism dramatically increases $k_\mathrm{on}^\mathrm{eff}$ and thereby reduces the effective dissociation constant typically by more than one order of magnitude, e.g. for XMAP215 we find $K^\tn{eff}\approx10^{-2} K$.
In the case of long dwell times $\omega_\tn{off}^{-1}$ and fast diffusion $\epsilon$, $K^\tn{eff}$ reduces to a particularly simple form 
\begin{equation}
K^\tn{eff} = (k_\tn{off} / \omega_\tn{on}) / \sqrt{\epsilon/\omega_\tn{off}} \, ,
\end{equation} 
where the denominator is the square root of the average number of diffusive steps a protein performs on the filament. Note that one-dimensional diffusion without capturing~\cite{Klein2005} does not lead to a particle flux on the filament [Supporting Material Fig.~\ref{fig:comparison}] and hence the reaction velocity is not increased [Fig.~\ref{fig:velocity}A]. 
Further, the particle flux might be limited by the length of the filament: Below a threshold length $l^c$ (which is smaller than typical \emph{in vivo} lengths of MTs) we observe a length dependent behavior of the reaction velocity [Supporting Material  Fig.~\ref{fig:length_dependency}] where our theory is not valid. 

Our analysis reveals \emph{diffusion and capture} as an efficient mechanism to circumvent the diffusion limit for the rate of end-targeting: Smoluchowski's theory of three-dimensional diffusion physically limits the rate of direct tip-attachment from solution~\cite{Smoluchowski1917}. As shown here, one-dimensional diffusion along a filament and subsequent capture at the filament end overcomes this limitation. This has been shown experimentally for MCAK~\cite{Helenius2006}. 
Our work provides an applicable theory for reaction kinetics facilitated by diffusion and capture: specific parameter values for diffusion, tip-association and dwell times can be accounted for, cf. Eqs.~\ref{eq:kondiff} and \ref{eq:speed}. On a broader perspective our results may also be applicable to other systems where one-dimensional diffusion is important~\cite{Hippel1989} including transcription factor binding on DNA~\cite{Hammar2012}.

\vspace{6mm}
\noindent\textbf{Acknowledgement}

{The authors thank Scott Hansen and Dyche Mullins for helpful correspondence on diffusing actin binding proteins.
This research was supported by the Deutsche Forschungsgemeinschaft (DFG) via project B02 within the SFB~863.}

\end{bibunit}

\definecolor{darkgreen}{rgb}{0,0.5,0} 
\definecolor{violet}{rgb}{0.5,0,0.5}
\definecolor{orange}{rgb}{0.2,0.5,0.5}
\newcommand{\bulk}[0]{\text{b}} 
\newcommand{\res}[0]{-} 

\newcommand{\bequ}{\begin{equation}}
\newcommand{\eequ}{\end{equation}}
\newcommand{\bequa}{\begin{eqnarray}}
\newcommand{\eequa}{\end{eqnarray}}
\newcommand{\bse}{\begin{subequations}}
\newcommand{\ese}{\end{subequations}}

\renewcommand{\theequation}{S\arabic{equation}}
\renewcommand{\thefigure}{S\arabic{figure}}
\renewcommand{\figurename}{FIGURE}

\renewcommand{\thesuppfigure}{S\arabic{suppfigure}}
\renewcommand{\thetable}{S\arabic{table}}
\renewcommand{\tablename}{TABLE}
\setcounter{equation}{0}
\renewcommand{\theequation}{S\arabic{equation}}
\newpage
\appendix
\begin{bibunit}
\begin{center}

\textbf{\huge{Supporting Material ``Quantifying protein diffusion and capture on filaments''}} 
\vspace{12pt}

\text{Emanuel Reithmann, Louis Reese, and Erwin Frey (\href{mailto:frey@lmu.de}{\nolinkurl{frey@lmu.de}})}
\vspace{12pt}

\textit{Arnold Sommerfeld Center for Theoretical Physics (ASC) and Center for NanoScience (CeNS), Department of Physics, Ludwig-Maximilians-Universit\"{a}t M\"{u}nchen, Theresienstra\ss e 37, D-80333 Munich, Germany, and Nanosystems Initiative Munich (NIM), Ludwig-Maximilians-Universit\"{a}t M\"{u}nchen, Schellingstra{\ss}e 4, D-80333 Munich, Germany}

\end{center}

\section{Derivation of Eq. (1), (2), and (4)}
In the following we derive the reaction velocity of the model presented in the Main Text. 
We start by considering the model depicted in Fig.~\ref{fig:Smodel} (A): Particles from a reservoir with concentration $c$ can bind reversibly to a single reaction site with rates $k_\mathrm{on} c$ for binding and $k_\mathrm{off}$ for unbinding. As long as the proteins are bound to the tip, they perform a not specified reaction at rate $k_\mathrm{cat}$. 
\begin{suppfigure}[h!]
\centering{\includegraphics[width=20pc]{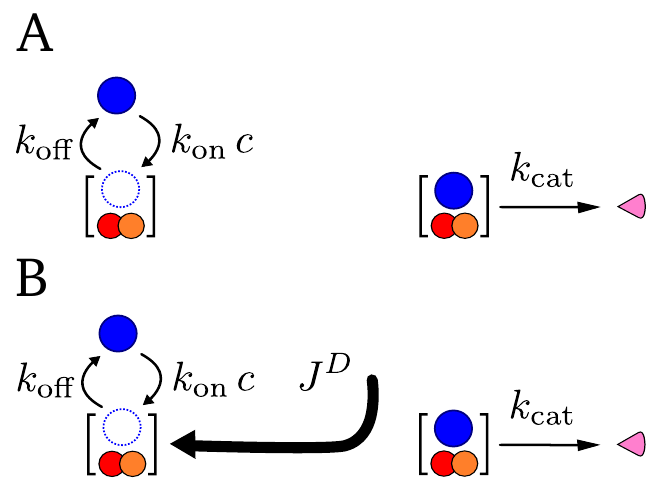}}
\caption{Illustration of the chemical reaction considered in this section. Particles (blue circles) from a reservoir with concentration $c$ bind reversibly to an unoccupied (dashed circles) reaction site at rates $k_\mathrm{on} c$ for binding and $k_\mathrm{off}$ for unbinding. Whilst bound, the particles catalyze a (not specified) reaction at rate $k_\mathrm{cat}$. In (A) particles can only attach directly via the reservoir. In (B) we have an additional particle flux $J^D$ to the reaction site due to a diffusion and capture mechanism described in the Main Text.
\label{fig:Smodel}}
\end{suppfigure}
Inspired by the processive (de)polymerase activity of XMAP215 (MCAK), there is no decomposition of the particle and the reaction site in the reaction step. Including a decomposition in the model would however be straightforward (see also the discussion in the Main Text). Let $n_+$ denote an occupied ($n_+=1$) or vacant ($n_+=0$) reaction site. Then, the average velocity of the reaction is given by
\begin{equation}
  v = \avg{n_+} k_\mathrm{cat}.
\end{equation}
Here, the average refers to an ensemble average. Since the reaction site corresponds to the last lattice site of one protofilament, we expect our results to be valid for experimental setups with a sufficiently large and constant number of protofilaments or (due to ergodicity) for the time average with respect to a single protofilament. In the steady state the equation for the average occupation of the reaction site reads
\begin{equation}
  0=\frac{d}{dt}\avg{n_+}= k_\mathrm{on} c (1-\avg{n_+}) - k_\mathrm{off} \avg{n_+} \, .
\end{equation}
Solving for $\avg{n_+}$ leads to an equation for the reaction velocity which is analogous to a single molecule Michaelis-Menten equation~\cite{Kou2005}:
\begin{equation}
\label{Eq:MM}
  v = \frac{k_\mathrm{cat} c}{K+c}
\end{equation}
with $K = k_\mathrm{off} / k_\mathrm{on}$. Note that here $c$ is the concentration of the enzyme, not the substrate.

If we now have an additional particle flux $J$ to the reaction site via one-dimensional diffusion, Fig.~\ref{fig:Smodel} (B), the dynamics are changed. Including this flux in the equation for the average reaction site occupation leads to
\begin{eqnarray}
  0=\frac{d}{dt}\avg{n_+}&=& k_\mathrm{on} c  (1-\avg{n_+}) + J - k_\mathrm{off}\avg{n_+}  \\
  &=& \left(k_\mathrm{on} c + \frac{J}{1-\avg{n_+}}\right) (1-\avg{n_+}) - k_\mathrm{off}\avg{n_+} \, .
\end{eqnarray}
Note that the term $J^\mathrm{D}:=J/(1-\avg{n_+})$ can be interpreted as the conditional particle flux to an unoccupied tip.
If the conditional diffusive current can be written as $J^\mathrm{D}:= k_\mathrm{on}^D c$ we conserve the functional form of Eq.~\ref{Eq:MM} but $k_\mathrm{on}$ is replaced by an effective on-rate for direct binding as well as diffusion and capture: $k^\mathrm{eff}= k_\mathrm{on} + k_\mathrm{on}^D$.

In conclusion, observing a Michaelis-Menten curve for the reaction velocity in dependence of the enzyme concentration $c$ is equivalent to the statement, that the current of particles towards an unoccupied reaction site due to the diffusion and capture mechanism obeys first order kinetics. In Fig.~\ref{fig:linearity} simulation results of our model show that the linearity condition is indeed approximately fulfilled over a broad parameter range. The corresponding parameter values are chosen within the typical parameter range for one-dimensionally diffusing proteins (see Fig.~\ref{fig:class} in the Main Text).
\begin{suppfigure}[h!]
\centering{\includegraphics[width=40pc]{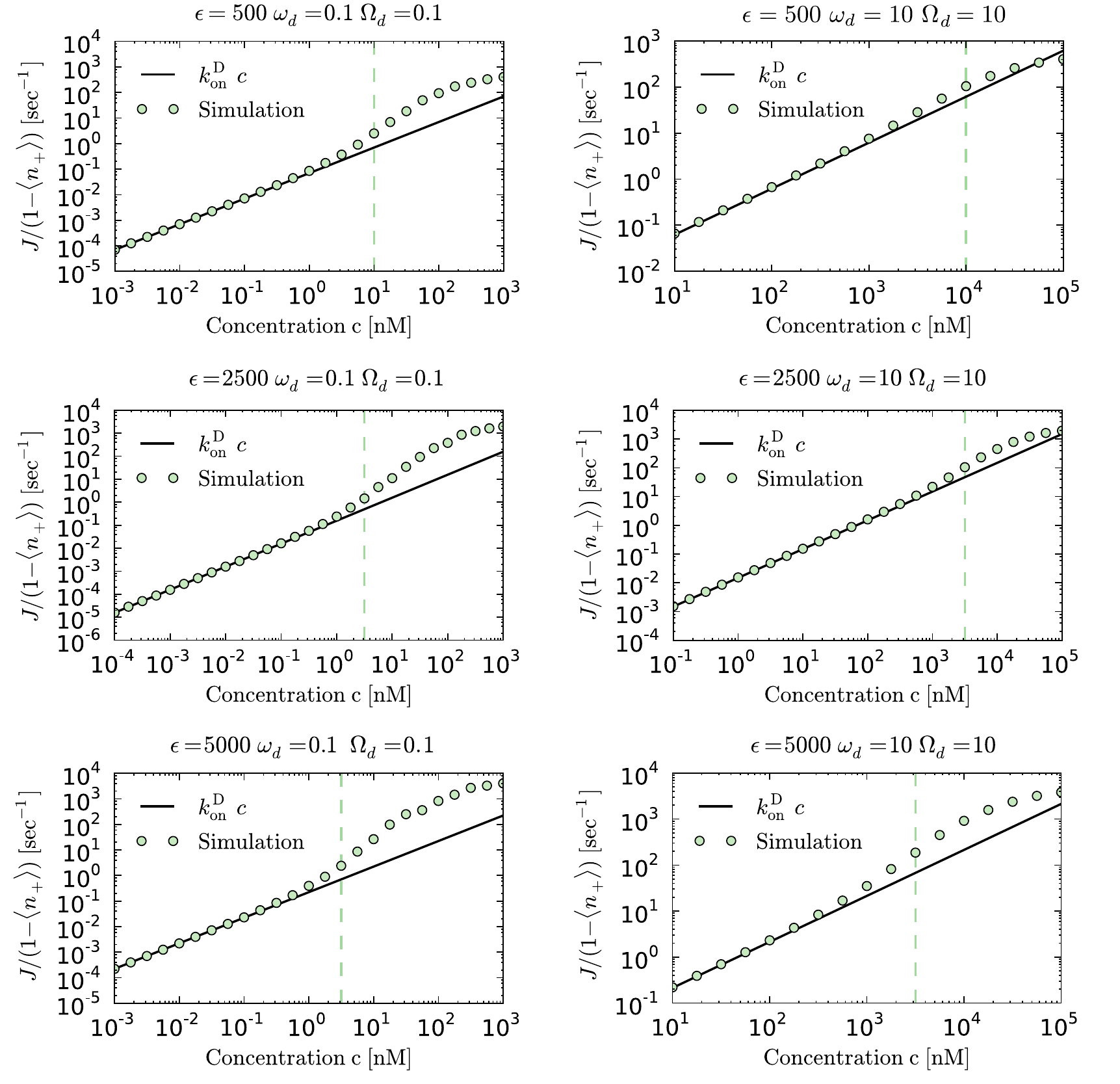}}
\caption{The current due to the diffusion and capture mechanism towards an unoccupied reaction site shows linearity in $c$ over a broad parameter range. The dashed line is the concentration, where the reaction velocity is 90\% saturated, $v=0.9\ k_\mathrm{cat}$. For concentrations close to the maximum reaction velocity the current differs from the linear behavior. This is, however, largely negligible because $v$ and $J$ are almost saturated.
\label{fig:linearity}}
\end{suppfigure}
Note that for high concentrations which imply almost saturated reaction velocities we observe a, in general, non-linear behavior. In such a parameter range extended analytic approaches than the ones presented here would be necessary. For the analysis shown in this work these deviations are, however, not relevant as we are interested in quantities which are almost saturated at such concentrations (tip-occupation and reaction velocity; particle current to the tip).

\section{Derivation of Eq. (3)}
\subsection{Mathematical model definition}
The mathematical formulation of the model depicted in Fig.~\ref{fig:model} of the Main Text relies on a probabilistic description of the lattice site occupations. A basic introduction to lattice gases and related problems can be found in Ref.~\cite{Krapivsky2010}. The occupation numbers $n_i$ with $i\in \{0,\dots,L-1\}$ describe the configuration of particles on the lattice, where $n_i=0$ and $n_i=1$ stand for an empty or occupied lattice site respectively. 
The equations of motion for the density on each lattice site are obtained in terms of the mean occupation numbers $\avg{n_i}=\rho_i$. Further it is assumed that neighboring lattice sites are occupied independently, which is a mean-field approximation that reads $\avg{n_i n_{i+1}}=\avg{n_i}\avg{n_{i+1}}$. We will justify later on that this approximation is not a restriction for the computations performed to derive $k_\mathrm{on}^D$.
Given these preliminaries, the equations of motion in the bulk (sites $i=2, \dots, L-1$) of the lattice read~\cite{Klein2005}
\begin{equation}
\label{Eq:RhoBulk}
\frac{d}{dt}\rho_i= \epsilon (\rho_{i-1}-2\rho_i + \rho_{i+1})+ \omega_\mathrm{on} c (1-\rho_i) - \omega_\mathrm{off}\rho_i \, ,
\end{equation}
where particle exclusion as well as attachment and detachment kinetics are accounted for. 
Particle attachment and detachment define the equilibrium density of particles on the lattice, also called the Langmuir density, 
\begin{equation}
\rho_\text{La}=\frac{\omega_\mathrm{on} c }{\omega_\mathrm{on} c + \omega_\mathrm{off}}\, .
\end{equation}
Note that we consider the terminal site separately and refer to its average occupation as tip density $\rho_+$,
\begin{equation}
\rho_+:=\rho_0=\avg{n_0} \, .
\end{equation}
The lattice site next to the tip, $i=1$, is also considered distinct from the bulk dynamics: It serves as a boundary for the diffusive region. The equations of motion for sites $i=0,1$ are
\begin{eqnarray}
\frac{d}{dt}\rho_0&=&\epsilon(1-\rho_0)\rho_1 + k_\mathrm{on} c (1-\rho_0)- k_\mathrm{off} \rho_0 \, \label{rho_nought},\\
\frac{d}{dt}\rho_1&=&\epsilon(\rho_2-\rho_1) -\epsilon \rho_1 (1-\rho_0) + \omega_\mathrm{on} c (1-\rho_1)- \omega_\mathrm{off} \rho_1 \, . \label{rho_one}
\end{eqnarray}
The density in the bulk is (in leading order of a gradient expansion) governed by the following diffusion equation
\begin{equation}
\partial_t \rho(x,t)=a^2  \epsilon \, \partial^2_x \rho(x,t) + \omega_\mathrm{on} c (1-\rho(x,t)) - \omega_\mathrm{off} \rho(x,t)\, ,
\end{equation}
with $x=a i$. This is a continuous approximation of Eq.~\ref{Eq:RhoBulk}. 
The boundary conditions are $\lim_{x \to \infty} \rho(x) = \rho_\mathrm{La}$ and $\rho(a) = \rho_1$. The solution is obtained in the steady state $\partial_t \rho(x,t)=0$ and reads
\begin{equation}
\rho(x)=\rho_\text{La} + (\rho_1-\rho_\text{La}) \exp[-(x-a)/\lambda]\, \label{rho_bulk_x},
\end{equation}
with
\begin{equation}
\lambda=\sqrt{\frac{D}{\omega_\mathrm{on} c + \omega_\mathrm{off}}}\, ,
\end{equation}
as calculated by Klein et al.~\cite{Klein2005}. Here we used $D=\epsilon a^2$.
This solution is valid for $x\geq a$. At sites $i=0$ and $i=1$ the density profile is not continuous and the diffusion equation can not be applied.
The remaining task is to determine the values for $\rho_0$ and $\rho_1$ in the steady state.

\subsection{Low density approximation}
To make progress, the particle flux from site $i=2$ to site $i=1$ from Eq.~\ref{rho_one} is rewritten as a derivative: $(\rho_2-\rho_1)\approx  a \partial_x \rho(x)|_{x=a}$. This corresponds again to a continuous description at the corresponding sites.
Using Eq.~\ref{rho_bulk_x} we see that the (right) derivative at $x=a $ (which is equivalent to $i=1$) is $\partial_x \rho|_{x=a+} = -(\rho_1-\rho_\text{La})/\lambda$. With this result we can simplify Eq.~\ref{rho_one} in the steady state
\begin{equation}
0\approx- a \frac{\rho_1 - \rho_\text{La}}{\lambda} - \rho_1 (1-\rho_0)\,\label{Eq:BC} ,
\end{equation}
where attachment and detachment rates have been assumed small. 
Further, note that we solve Eq.~\ref{rho_nought} in the steady state for $\rho_0$ as a function of $\rho_1$:
\begin{equation}
\rho_0 = \frac{\rho_1 \epsilon  + k_\text{on} c}{\rho_1 \epsilon + k_\text{on}c + k_\text{off}} \, \label{Eq:rhoPlus}  .
\end{equation}
The key relation at the basis of our theory is that tip attachment via lattice diffusion obeys first-order reaction kinetics. This is well confirmed by simulations over a broad and biologically relevant parameter region, see Fig.~\ref{fig:linearity}. The current to an unoccupied tip is hence approximately given by $J^D = k_\mathrm{on}^D c$, where $k^\text{D}_\text{on}$ is a constant, independent of the enzyme concentration. In the following, we determine the diffusive current for infinitesimally low enzyme concentrations and thereby determine $k^\text{D}_\text{on}$. In this parameter region correlations become negligible such that the mean-field assumption becomes valid.
Further, we assume $\rho_1$ to be small. This is well justified for low concentrations. Note also that the density along the lattice is minimal at $i=1$. Up to first order in $\rho_1$ Eq.~\ref{Eq:BC} reduces to
\begin{equation}
0 = - \frac{a\,  \epsilon \, \omega_\text{on} c}{\sqrt{ \epsilon \, (\omega_\text{off}+\omega_\text{on} c)}} + \bigl(a\, \sqrt{ \epsilon \,  (\omega_\text{off}+\omega_\text{on} c)} + \frac{ \epsilon \, k_\text{off}}{k_\text{on} c + k_\text{off}}\bigr) \rho_1\, .
\end{equation}
The solution of the above equation determines the tip density via Eq.~\ref{Eq:rhoPlus}. In our low-density approximation and up to first order in $c$ we obtain
\begin{equation}
\rho_0^\text{low-c}= \frac{ k_\text{on} +( \omega_\mathrm{on} \, \epsilon) /(\omega_\text{off}+\sqrt{ \epsilon \, \omega_\text{off}})}{k_\text{off}} \ c \label{Eq:rhoPlusSol} \, .
\end{equation}

\subsection{Site attachment due to lattice diffusion}
The solution of $\rho_+ = \rho_0$ allows us to determine the diffusive current. In Eq.~\ref{Eq:rhoPlusSol} there is an additional term that adds to the direct attachment rate $k_\text{on}$ which vanishes for $ \epsilon =0$. Identifying this term as the diffusive on-rate the final result reads
\begin{equation}
k_\text{on}^\text{D}=\frac{ \epsilon  \, \omega_\text{on}}{\omega_\text{off}+\sqrt{ \epsilon \,  \omega_\text{off}}} \,.
\end{equation}

\section{XMAP215 parameter values}
\begin{table}[h!!]
\setlength{\extrarowheight}{2.pt}
\begin{tabular*}{0.95\textwidth}{@{\extracolsep{\fill}}ccccccc}
\hline \hline
\multirow{2}{*}{Model parameters} & $\epsilon $ & $\omega_\mathrm{on}$ & $\omega_\mathrm{off}$ & $k_\mathrm{cat}$ & $k_\mathrm{off}$ & $k_\mathrm{on}$ \\ 
& $\mathrm{\ s}^{-1}$ & $\mathrm{\ (nM\ s)}^{-1}$ & $\mathrm{\ s}^{-1}$ & $\mathrm{\ s}^{-1}$ & $\mathrm{\ s}^{-1}$ & $\mathrm{\ (nM\ s)}^{-1}$  \\ \hline
XMAP215 & 4.7 $\times 10^{3}$\ & 6 $\times 10^{-5}$ & 4.1 $\times 10^{-1}$ & 1.3 $\times 10^2$ & 2.6 $\times 10^{-1}$  & 6 $\times 10^{-5}$\\\hline\hline
 \end{tabular*}
  \caption{Parameters used in our simulation for XMAP215. The numbers were derived from experimental data~\cite{Brouhard2008,Widlund2011}.} 
\label{tab:MCAK_rates}
\label{tab:MCAK_exp}
\end{table}

\FloatBarrier
\section{Length dependent behavior for short filaments}
For filaments shorter than a critical length $l_c$ our model implies a length dependent tip-density, see  Fig.~\ref{fig:length_dependency}. Below $l_c$ the particle flux due to diffusion and capture is limited because the filament is shorter than the length scale of free diffusion. For XMAP215 we find $l^c \approx 0.5\, \mu\mathrm{m}$. This means that for typical \emph{in vivo} lengths of MTs, where $l>l^c$, the enhancement of tip association through one-dimensional diffusion and capture is realized. However, below $l^c$ length dependent behavior might occur.

\begin{suppfigure}[h!!]
\centering{\includegraphics[width=20pc]{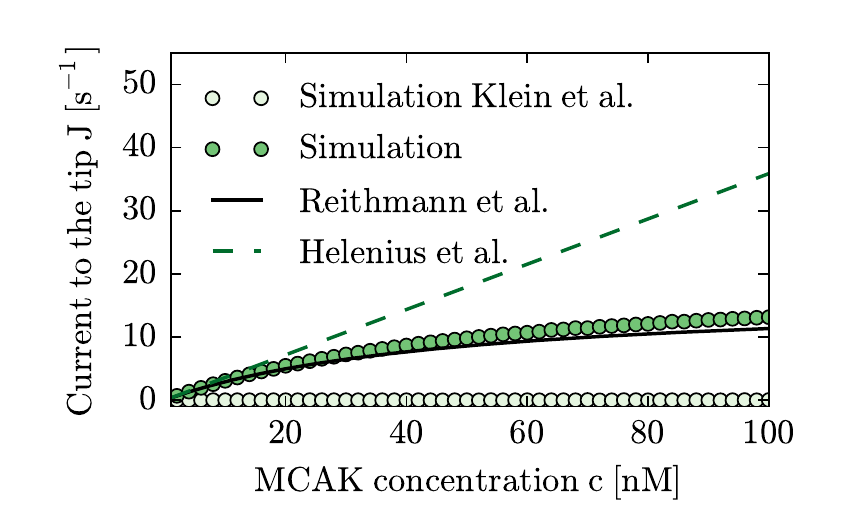}}
\caption{Below a critical length $l^c$ (dotted) the reaction velocity $v$ depends on the MT length. The reaction velocity saturates at the value given by Eq. 4 in the Main Text. Parameters are for XMAP215, $c=300\ \mathrm{nM}$, see Table~\ref{tab:MCAK_exp}.
\label{fig:length_dependency}}
\end{suppfigure}

\FloatBarrier
\section{Differences to work by Helenius et al.~\cite{Helenius2006} and Klein et al.~\cite{Klein2005}}

There are several differences between our work and previous theoretical investigations of the topic. 
Work by Helenius et al.~\cite{Helenius2006} did not address the question of reaction kinetics and motor occupations at the MT tip. 
Their approach based on differential equations leads to mathematical inconsistencies when considering a capturing mechanism.
This is why we pursued a different approach using a lattice gas. 
It allows us to account for the capturing mechanism which in turn leads to an explicit expression for the tip density. 
If the tip is highly occupied, the current saturates at a value which is approximately given by the off-rate at the tip, $k_\tn{off}$. This saturation effect is not included in the model by Helenius et al.~\cite{Helenius2006}, but is accounted for in our model.
In Fig.~\ref{fig:comparison} we compare the particle current $J$ obtained by Helenius et al. with our result. For high concentrations, there are large deviations between our model and the approximation in~\cite{Helenius2006}.
Work by Klein et al.~\cite{Klein2005} did not include capturing at the MT tip. 
Therefore there is a vanishing current to the tip if diffusion is fast compared to the enzymatic reaction at the tip, cf. Fig.~\ref{fig:comparison}.

\begin{suppfigure}[h!]
\centering{\includegraphics[width=23pc]{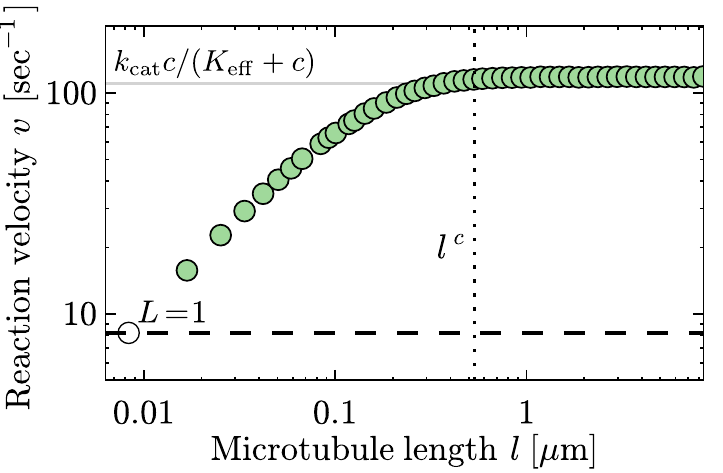}}
\caption{Differences in the (total) protein current $J$ to the MT tip between this work and previous theoretical approaches. A continuous diffusion equation with absorbing boundary condition leads to a linear relation $J\propto c$~\cite{Helenius2006}. A lattice gas without capturing~\cite{Klein2005} (reflecting boundary condition) has a vanishing current to the tip $J =0$. 
\label{fig:comparison}}
\end{suppfigure}

\FloatBarrier

\end{bibunit}

\end{document}